\documentclass{PoS}
\usepackage{graphicx}
\usepackage{subfigure}

\title{Scaling, topological tunneling and actions for weak coupling DWF calculations}

\ShortTitle{Scaling, topological tunneling and actions for weak coupling DWF calculations}

\author{\speaker{Greg McGlynn}, Robert D. Mawhinney\\
        Physics Department, Columbia University, New York, NY 10027, USA\\
        E-mail: \email{gem2128@columbia.edu}, \email{rdm@physics.columbia.edu}}

\abstract{We present results from a 2+1 flavor DWF calculation at $a^{-1} = 3$
GeV and discuss strategies for similar calculations at finer lattice spacings
which will target charm physics. At weak coupling the autocorrelation time of
the global topological charge becomes very long because the HMC algorithm has
trouble moving between topological sectors. We report the results of
simulations that test two ideas for reducing the autocorrelation time of
topological charge. In weak coupling quenched simulations we find that the open
boundary conditions suggested by L\"{u}scher and Schaefer do not prevent the
appearance of extremely long autocorrelation times for topological observables.
We discuss the idea of a ``dislocation-enhancing determinant'' and show that it
can produce an increase in topological tunneling.}

\FullConference{31st International Symposium on Lattice Field Theory LATTICE 2013\\
                 July 29 . August 3, 2013\\
                 Mainz, Germany}

\begin{document}

\section{Motivation}

Long autocorrelations in Monte Carlo simulations make it hard to reliably
estimate statistical errors on measured quanitites. In lattice QCD simulations,
the autocorrelation time of the topological charge increases alarming as the
lattice spacing is reduced \cite{Alpha}. Depending on the gauge action, at
lattice spacings of order 0.05 fm and below the autocorrelation time of the
topological charge can be comparable to the typical length of a lattice QCD
simulation.  

\begin{figure} \centering
\mbox{
  \subfigure[Topological charge]{\includegraphics[width=2.7in]{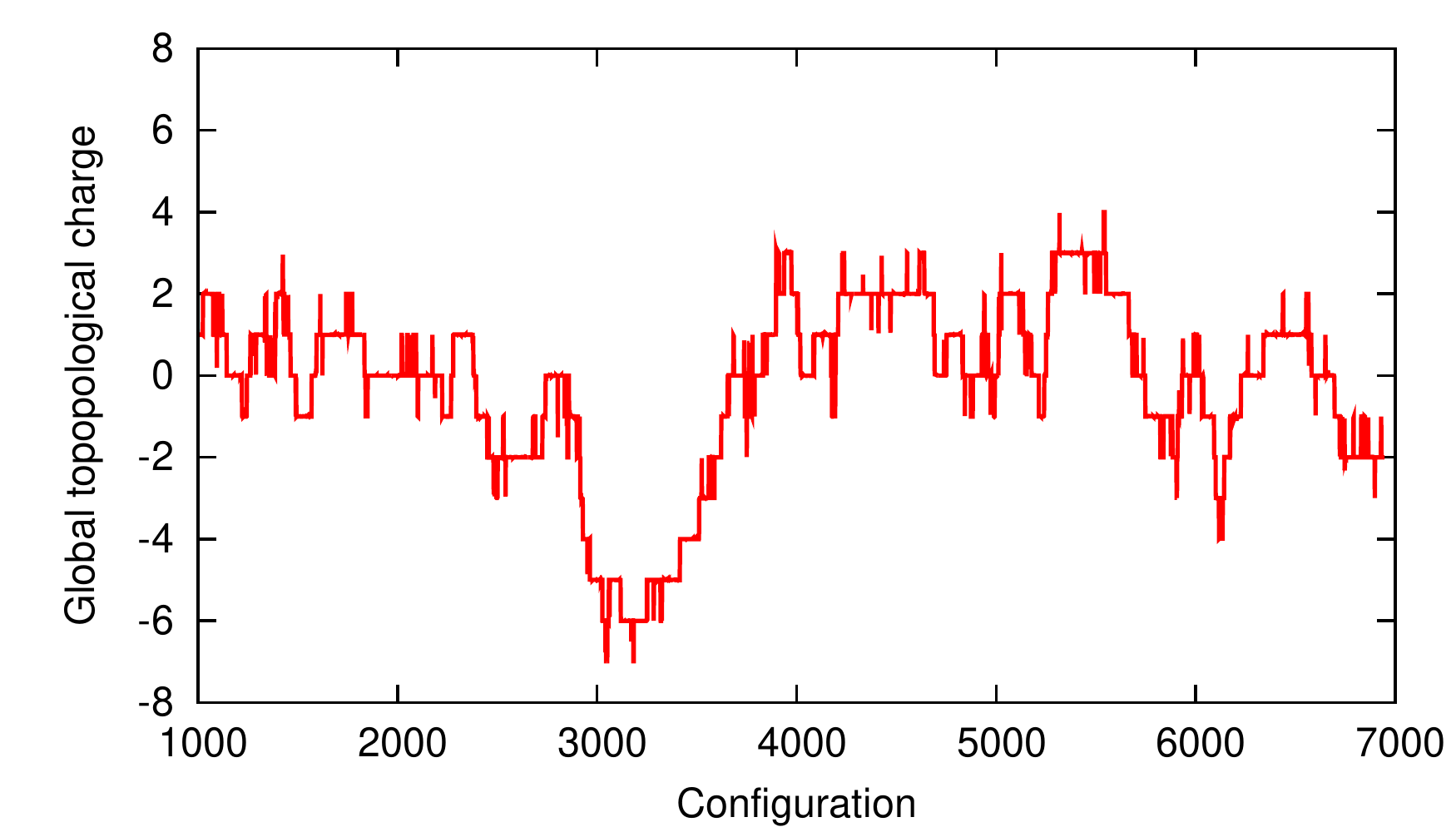}} \quad
  \subfigure[$\eta_c$ dispersion relation]{\includegraphics[width=2.7in]{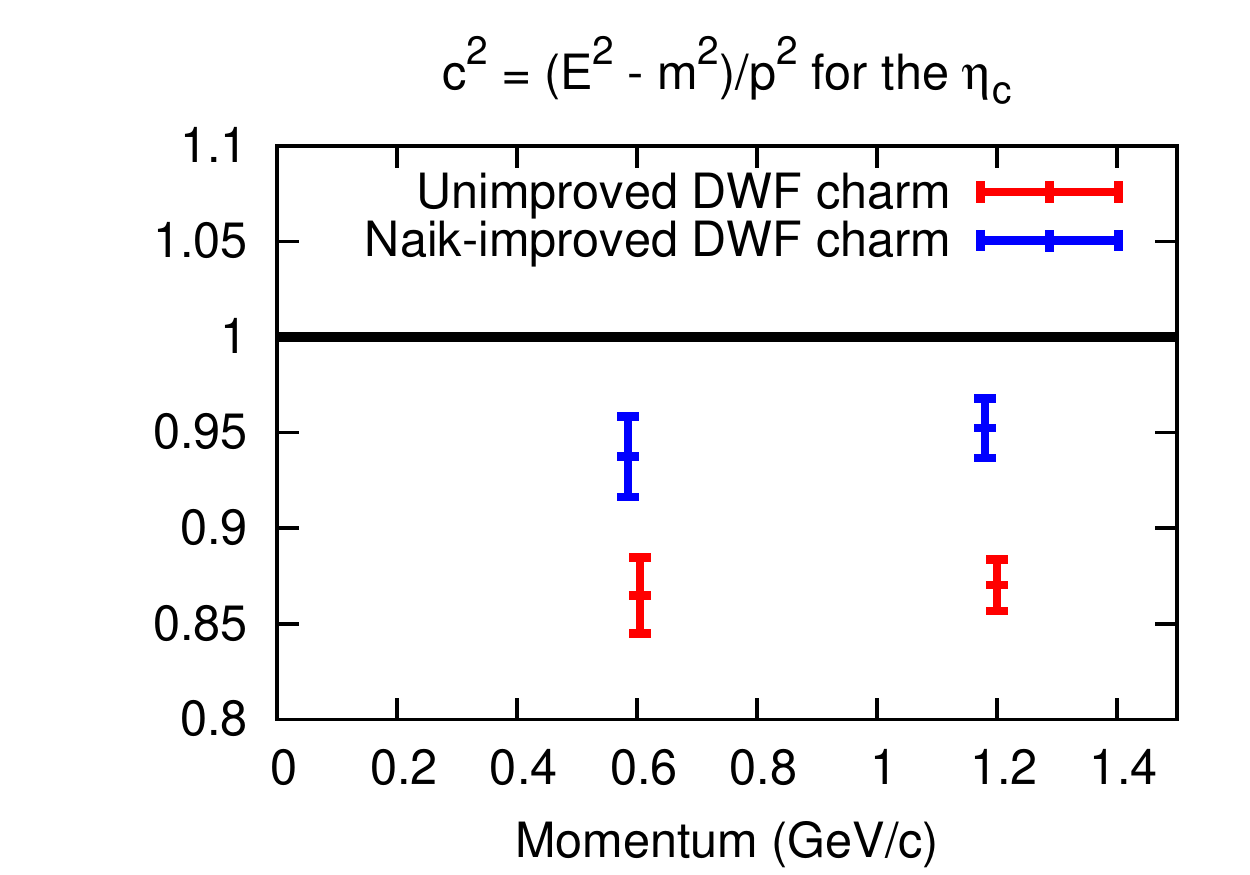}}
}
\caption{Measurements on a 2+1 flavor lattice QCD simulation with $a =
0.065(2)$ fm and $m_\pi \approx 360$ MeV.}
\label{fig:3GeVsim} 
\end{figure}

As an example of the problem with topology, Figure \ref{fig:3GeVsim}a shows the
topological charge $Q$ as a function of MD time for a QCD simulation with 2+1
flavors of domain-wall quarks at $a=0.065(2)$ fm. The integrated
autocorrelation time (IAT) of $Q$ is of order 250 MD time units. While this
might be (barely) tolerable, the autocorrelation time is expected to grow very
quickly as $a$ decreases, such that the IAT will soon be comparable to or
greater than the length of the simulation itself. For example at $a=0.05$ fm
the autocorrelation time of $Q$ would probably be unacceptably long and we
would have to worry about whether statistical errors could be estimated
reliably. 

But we would very much like to simulate at $a \lesssim 0.05$ fm in order to
reduce the discretization errors associated with physical-mass charm quarks.
Figure \ref{fig:3GeVsim}b shows $c^2 \equiv (E^2 - m^2)/p^2$ for the
physical-mass $\eta_c$ meson measured on the $a=0.065$ fm ensemble with
domain-wall valence charm quarks. The $\sim 13\%$ difference from the correct
value $c^2 = 1$ is a measure of $O(4)$ symmetry breaking due to lattice
artifacts. This is reduced to $\sim 7\%$ with Naik improvement, and going to $a
= 0.05$ fm would bring these $O(a^2)$ discretization errors below 5\%, which
would be very helpful in taking the continuum limit. Another motivation for
decreasing $a$ is that finer lattice spacings will also allow for better
matching of lattice simulations to perturbation theory by allowing the matching
to be performed at a higher energy scale where perturbation theory is more
reliable.

Therefore it would be very useful to find some way to reduce the
autocorrelation time of the topological charge in weak coupling lattice QCD
simulations. We discuss results from experiments to test two ideas for reducing
the autocorrelation time of $Q$: open boundary conditions and a
``dislocation-enhancing determinant.''

\section{Open boundary conditions}

L\"{u}scher and Schaefer have proposed using open boundary conditions in the
time direction to reduce the autocorrelation time of $Q$ \cite{SLOBC}. With the
usual periodic boundary conditions, if the gauge field is smooth (as it is at
small $a$), $Q$ is quantized and so can only change by ``tunneling'' between
disconnected sectors of the field space. These tunneling events are rare at
small $a$. But if we choose open boundary conditions in at least one direction,
then $Q$ is no longer quantized and can change continuously as topological
charges flows in or out through the lattice boundaries, without the need for
tunneling events. This may let $Q$ change more quickly, reducing its
autocorrelation time.

We ran experiments with two different gauge actions to test whether open
boundary conditions indeed reduce the autocorrelation time of $Q$ at weak
coupling. We simulated pure SU(3) gauge theory with the HMC algorithm. (We are
ultimately interested in improving simulations with dynamical fermions, which
use the HMC algorithm, so our pure gauge theory experiments use HMC rather than,
say, a heat bath update algorithm). Table \ref{tab:OBCtable} describes the
parameters of each simulation. In each experiment we compared a reference
ensemble with periodic boundary conditions to an otherwise identical ensemble
with open boundary conditions in time direction.  

\begin{table} \centering
\begin{tabular}{c|c|c}
Gauge action        & Iwasaki                 & Wilson             \\ \hline \hline
$\beta$             & 2.9                     & 6.42               \\ \hline
$a$                 & 0.069(2) fm \cite{IwasakiSpacing}  & 0.0500(4) fm \cite{WilsonSpacing} \\ \hline
Lattice volume      & $24^3 \times 64$        & $32^3 \times 32$   \\ \hline
Physical volume     & $(1.7 \mbox{ fm})^3 \times (4.4 \mbox{ fm})$ & $(1.6 \mbox{ fm})^4$ \\ \hline
%Trajectory length   & 1 MD unit               & 1 MD unit          \\ \hline
%Steps per trajectory& 10                      & 10                 \\ \hline
%Integrator          & Force gradient          & Force gradient     \\ \hline
%Acceptance rate     & 70 \%                   & 84 \%              \\ \hline
MD time units       & 18000                   & 5300               \\ \hline
\end{tabular}
\caption{Parameters of the open boundary conditions experiments. The Wilson
ensemble parameters were chosen to coincide with one of the runs in
\cite{SLOBC}.}
\label{tab:OBCtable}
\end{table}

When open boundary conditions are used, the simulated physics is distorted in a
layer near the open boundaries. For instance, the mean plaquette is different
near the boundaries than in the bulk of the lattice (by ``bulk'' we mean the
interior region of the lattice far from the boundaries where the physics is
independent of the boundary conditions). We are careful to compare the open and
periodic lattices only in the bulk. The boundary region to be excluded has a
width that is determined by two main effects. First, the influence of the
boundary should be expected to penetrate a distance into the bulk of order
$1/m$ where $m$ is the mass of the lightest state in the theory. In the pure
gauge theory this is a glueball with a mass $m \sim 1$ GeV, so that $1/m$ is of
order 3-4 lattice spacings at our values of $a$. Second, if cooling algorithms
such as link smearing or the Wilson flow are used, they effectively average the
gauge field over some radius and thus extend the region that should be excluded
because of the influence of boundary effects.

Because of these boundary effects, we do not want to compare the global
topological charge $Q$ between the open and periodic lattices. The next best
thing is to construct a subvolume topological charge by summing the topological
charge density over a large interior region of the lattice:
\begin{equation} Q(t_1, t_2) \equiv \sum_{t_1 \le t < t_2} \sum_{\vec{x}}
\rho(\vec{x}, t) \end{equation}
where $t_1$ and $t_2$ must be chosen to exclude the boundary regions.  For the
topological charge density $\rho(\vec{x}, t)$ we use the 5Li definition from
\cite{5Li}, measured after 60 rounds of APE smearing with coefficient 0.45.  As
mentioned above, link smearing increases the size of the boundary region. Our
parameters produce an RMS smearing radius of order 6 lattice spacings
\cite{APEradius}.

\begin{figure} \centering
\mbox{
  \subfigure[Iwasaki, $\beta = 2.9$]{\includegraphics[width=2.7in]{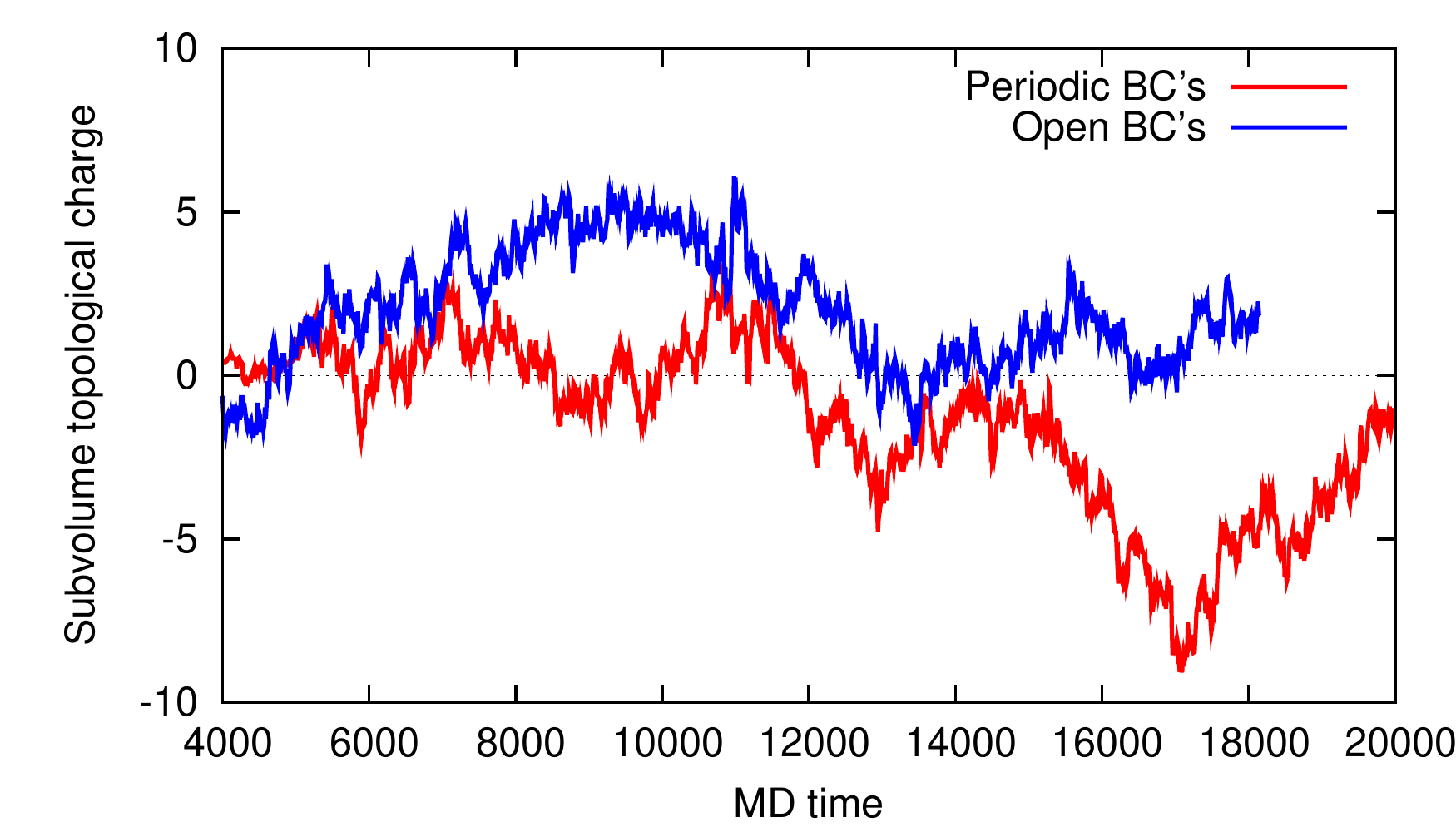}} \quad
  \subfigure[Wilson, $\beta = 6.42$]{\includegraphics[width=2.7in]{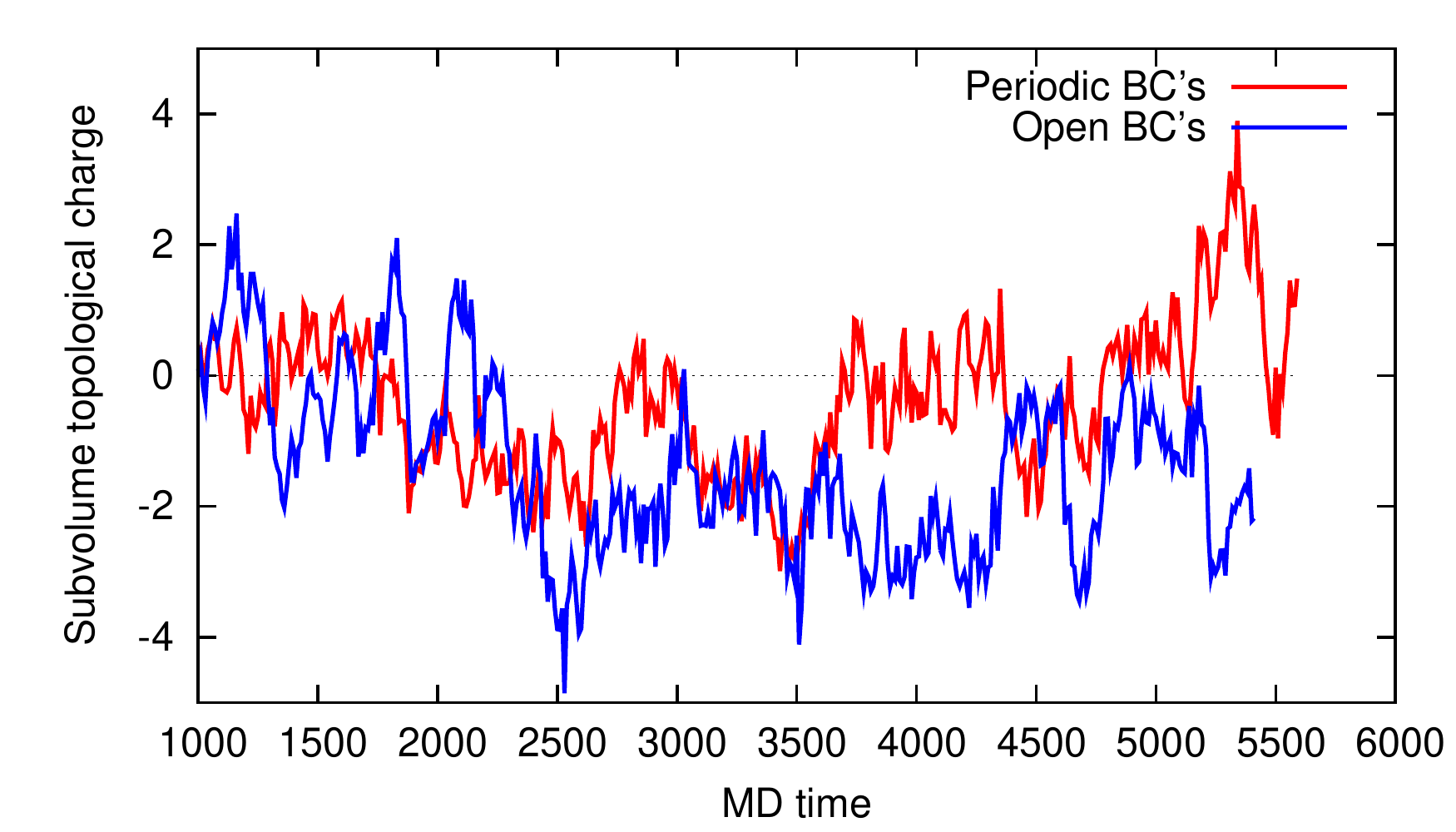}}
}
\caption{MD time history of the topological charge summed over large
subvolumes.  On the $24^3 \times 64$ Iwasaki ensembles we plot $Q(16, 48)$ (a
sum over the central 32 time slices), while on the $32^4$ Wilson ensembles we
plot $Q(8, 24)$ (a sum over the central 16 time slices).}
\label{fig:OBCsubvolumes} 
\end{figure}

For large subvolumes we observe very long autocorrelation times of the subvolume
charge, independent of the boundary conditions used. An example is shown in
Figure \ref{fig:OBCsubvolumes}. While the autocorrelation times are much too
long for us to be able to make reliable numerical estimates, they are clearly of
order thousands of MD time units on the Iwasaki ensembles and at least many
hundreds on the Wilson ensembles. Visually, the time histories for the open
lattices do not look any better than those for the periodic lattices.

\begin{figure} \centering
\mbox{
  \subfigure[Iwasaki, $\beta = 2.9$]{\includegraphics[width=2.7in]{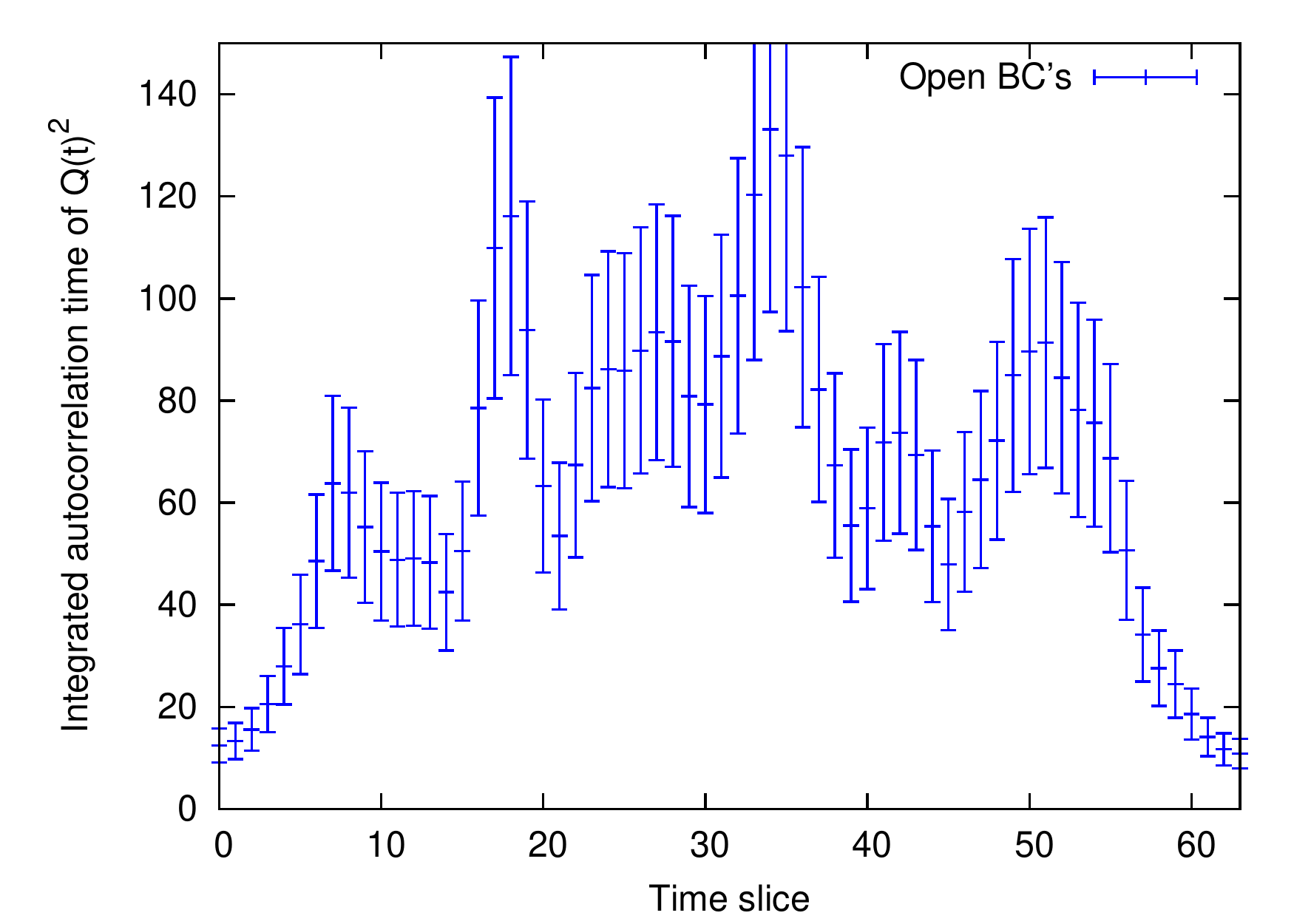}} \quad
  \subfigure[Wilson, $\beta = 6.42$]{\includegraphics[width=2.7in]{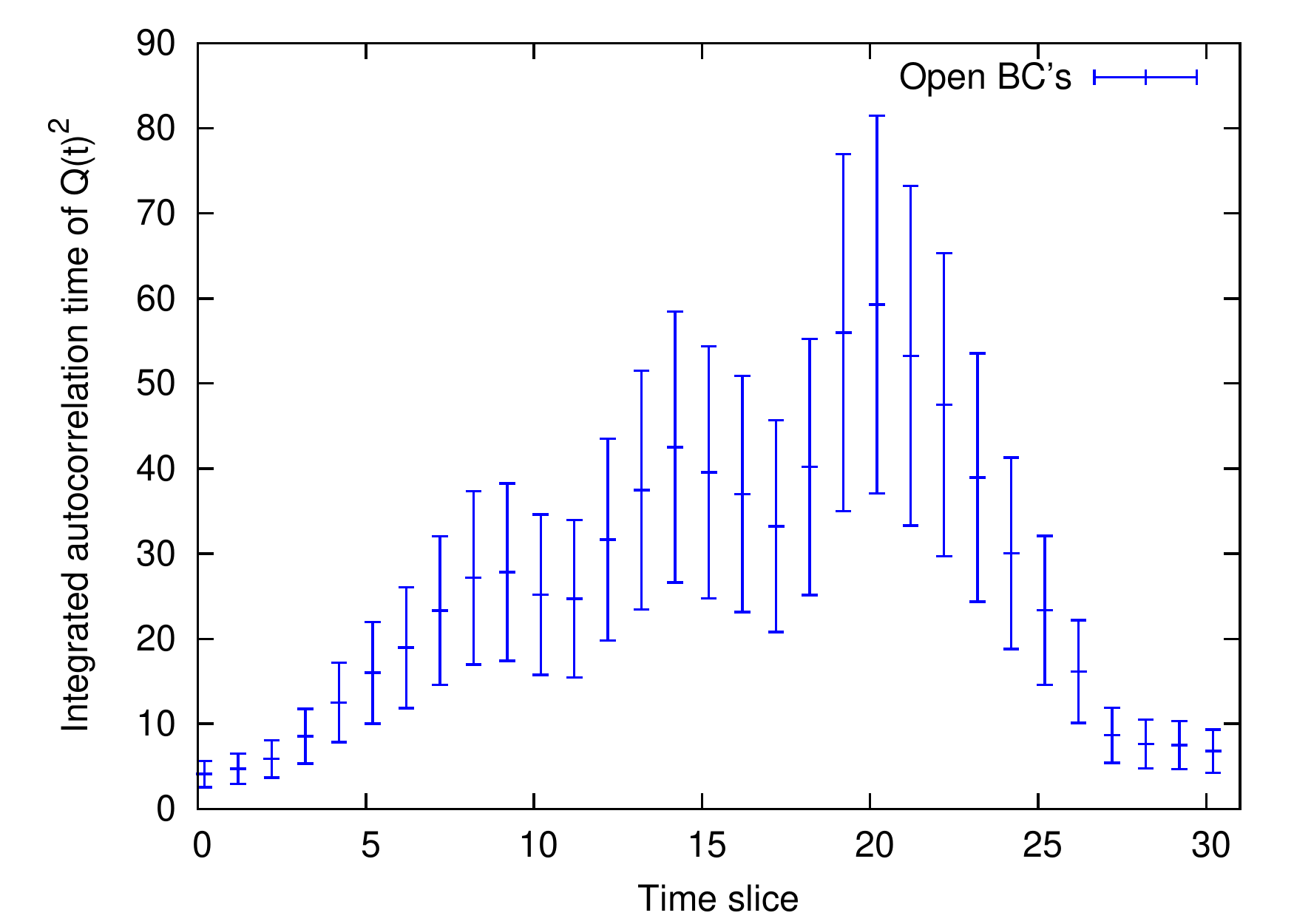}}
}
\caption{Measured integrated autocorrelation time of $Q(t)^2$ as a function of
$t$ on open lattices. The integration window was 250 MD time units on the
Iwasaki ensemble and 150 MD time units on the Wilson ensemble. Note that IATs
in the bulk are likely to be significantly underestimated beyond what is
suggested by the error bars because of the relatively short length of our
simulations.}
\label{fig:OBCsliceiats} 
\end{figure}

The open boundaries do have a dramatic effect on autocorrelation times within a
narrow region near the boundaries. In order to measure the effect of the open
boundaries as a function of the Euclidean time coordinate we can compute
$Q(t)\equiv{}Q(t,t+1)$, the topological charge summed over a single time slice.
Figure \ref{fig:OBCsliceiats} shows the integrated autocorrelation time of
$Q(t)^2$ as a function of $t$ on the open lattices. It is clear that open
boundary conditions reduce the integrated autocorrelation time of $Q(t)^2$
dramatically for time slices near the time boundaries compared to time slices
in the bulk. 

However what we really care about is improving autocorrelation times in the
interior region of the lattice, where the physics is independent of the
boundary conditions. When we measure $Q(t)^2$ on the periodic ensembles, we see
no difference between IATs on the periodic lattices and the IATs in the bulk
region of the corresponding open lattices. However, we caution that our runs
are far too short to estimate these integrated autocorrelation times reliably,
so that this comparison cannot be taken too seriously. The autocorrelation
function of $Q(t)^2$ likely has a long tail which we cannot measure well, so
that the integrated autocorrelation time is likely to be significantly
underestimated\footnote{We thank Martin L\"{u}scher and Stefan Schaefer for
useful discussions on this point.}.

Since we lack sufficient statistics to do a numerical comparison of
autocorrelation times, it is possible that open boundary conditions are
producing some improvement. But Figure \ref{fig:OBCsubvolumes} shows that even
if autocorrelation times in the interior region far from the boundary are
shorter with open boundary conditions, they are still unacceptably long.

\section{Dislocation-enhancing determinant}

Since a smooth gauge field cannot change its value of $Q$ continuously, lattice
simulations must ``tear'' the gauge field when they move between topological
sectors. These tears or ``dislocations'' are associated with zero eigenvalues
of the hermitian Wilson Dirac operator $H(M)$ for mass $M \sim -1/a$. For
example, each change in the index of the overlap Dirac operator is associated
with an eigenvalue of $H(M)$ passing through zero \cite{OverlapIndex}. (We
don't measure $Q$ from the index of the overlap Dirac operator, but all
definitions of $Q$ should be equivalent in the continuum limit.)

This suggests that we might increase the rate of transitions between
topological sectors by encouraging more zero-modes of $H(M)$ for $M\sim-1/a$.
The intuition is that there is an action barrier in the space of lattice gauge
fields between different topological sectors and that we can increase the rate
of tunneling through the barrier by decreasing the barrier height. In the
continuum limit dislocations will become vanishingly rare no matter what we do,
but at finite lattice spacing we can increase their density in the hope of
reaching smaller $a$ before topology freezes. To enhance dislocations we can
introduce an auxiliary determinant into the lattice action:
\begin{equation} \det f(H(M)) = \prod_i f(\lambda_i) \end{equation}
where $\{\lambda_i\}$ are the eigenvalues of $H(M)$. The nonnegative function
$f(\lambda)$ should go to $1$ for large $\lambda$ so that the actual QCD
physics is not affected. But the value of $f(\lambda)$ near $\lambda=0$ can be
adjusted to control the density of dislocations.

In the past people have usually instead tried to \emph{suppress} the zero-modes
of $H(M)$, for example because these modes increase the residual chiral
symmetry breaking in the domain-wall formulation of chiral fermions.  If $0
\leq f(0) < 1$ then dislocations are suppressed.  For example the RBC-UKQCD
collaboration has sucessfully used the ``dislocation-suppressing determinant
ratio'' 
\begin{equation} \label{eq:DSDR} f(\lambda) = \frac{\lambda^2 +
\epsilon_f^2}{\lambda^2 + \epsilon_b^2} \end{equation}
with $\epsilon_f \ll \epsilon_b$ to reduce the residual chiral symmetry
breaking in DWF simulations at strong coupling \cite{DSDR}. At weak coupling
the residual chiral symmetry breaking is already small and we can instead
contemplate making $f(0) > 1$ in order to enhance dislocations and speed up
topological tunneling. We call this a ``dislocation-enhancing determinant''
(DED).

Perhaps the most straightforward way of doing this is to use Eq.
(\ref{eq:DSDR}) with $\epsilon_f > \epsilon_b$. This ratio is easily simulated
using twisted-mass Wilson fermions. But we found that if we introduced a DED
using this form of $f$ we needed to increase $\beta$ substantially in order to
keep the lattice spacing unchanged. Increasing $\beta$ tends to suppress
topological tunneling and so we did not find much net improvement in the
tunneling rate from this form of $f$.

However there is a great deal of freedom to choose the form of $f$. We sought
an $f(\lambda)$ that went to zero very rapidly for large $\lambda$ in an
attempt to avoid affecting as many high modes of $H(M)$, which we expected
would reduce the $\beta$ shift needed to maintain the lattice spacing. We chose
\begin{equation} \label{eq:ratfun} f(\lambda) = \left(1-\frac{A}{\lambda^2+B_1}
+ \frac{A}{\lambda^2+B_2}\right)^2 \end{equation}
With this form $f(\lambda) = 1 + O(\lambda^{-4})$ for large $\lambda$, which is
a faster falloff than the form of Eq. (\ref{eq:DSDR}) where
$f(\lambda)=1+O(\lambda^{-2})$. This determinant is easily simulated using
existing RHMC codes. The constants $A$, $B_1$, and $B_2$ can be tuned to
control the zero-mode enhancement $f(0)$ and the width of the region around
zero in which eigenvalues are enhanced. For convenience in our simulations we
actually replace $H(M)$ with the hermitian even-odd preconditioned Wilson Dirac
operator
\begin{equation} H_{\rm prec}(M) = \gamma_5 \left((M+4) +
\frac{1}{M+4}D_{oe}D_{eo} \right) \end{equation}
where $D_{oe}$ and $D_{eo}$ are the parts of the Wilson Dirac operator that
connect sites of opposite parity. If $H$ has a zero eigenvalue then so does
$H_{\rm prec}$, so this still enhances dislocations.

\begin{figure} \centering
\includegraphics[width=3in]{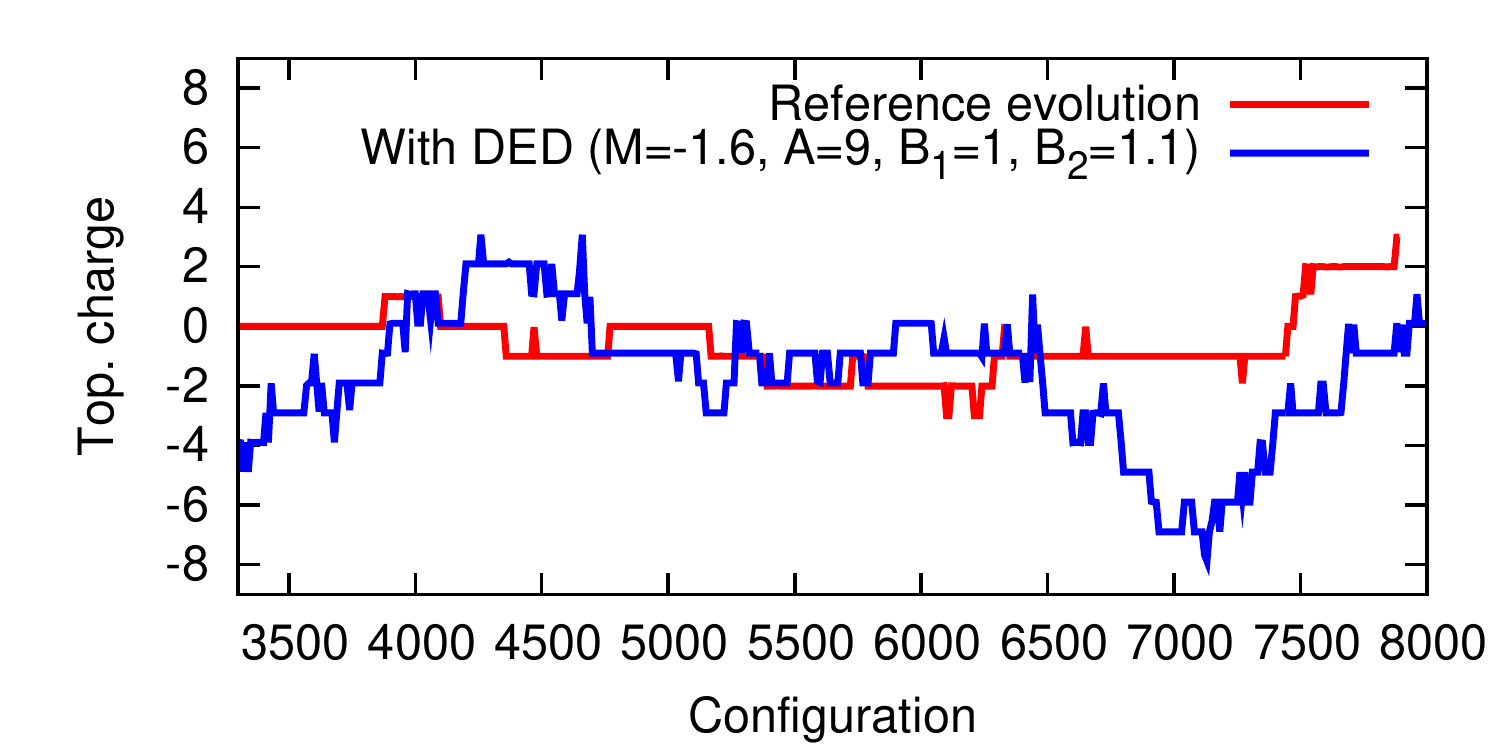}
\caption{Topological charge as a function of MD time measured on a DED ensemble
and a reference ensemble at $a=0.05$ fm.}
\label{fig:DEDtcharge} 
\end{figure}

Figure \ref{fig:DEDtcharge} shows the topological charge measured in a quenched
simulation using this DED with the Wilson gauge action at $a = 0.05$ fm. Also
shown for reference is the topological charge evolution on an ensemble at the
same lattice spacing using the regular unmodified Wilson gauge action. It is
important to make such comparisons at the same lattice spacing, since the
topological tunneling rate depends very sensitively on $a$. The lattice spacings
for the two ensembles were matched using the Wilson flow scale $t_0$
\cite{WilsonFlow} by adjusting the value of $\beta$ on the DED ensemble.  This
required increasing $\beta$ to 6.70 on the DED ensemble compared to 6.42 on the
reference ensemble.

Although the problem of long autocorrelations has by no means been solved, the
DED clearly improves the topological tunneling rate in Figure
\ref{fig:DEDtcharge}. The frequency with with $Q$ changes is about 5 times
greater in the DED ensemble compared to the reference ensemble with the
unmodified Wilson gauge action. Furthermore, the DED ensemble explores a much
wider range of topological sectors in the same span of MD time.
It is possible that even greater gains could be achieved with a different form
of $f$, or by using a different value of $M$.

%There is a lot of freedom to vary the dislocation-enhancing determinant idea to
%try to improve topological tunneling further. For instance, many choices of the
%function $f$ are possible; our choice was largely dictated by what was easy to
%simulate with existing code. The negative mass $M$ in $H(M)$ is also a free
%parameter.  One idea we are currently pursuing in experiments is to introduce
%several DED's with different values of $M$. The idea is that we want to provide
%a continuous path by which zero-modes of $H(M)$ with $M \sim -1/a$ can evolve
%into zero-modes of $H(m_c)$, where $m_c$ is the (negative) critical Wilson
%mass. Zero-modes of $H(m_c)$ should correspond to physical-size instantons. By
%enhancing zero-modes of $H(M_i)$ for a number of $M_i$ between $\sim -1/a$ and
%$m_c$, we can provide a continuous path of reduced action by which dislocations
%can grow into physical-sized instantons.

\section{Conclusions}

We have investigated two techniques intended to reduce the autocorrelation time
of the topological charge in lattice QCD simulations at weak coupling. In our
experiments with open boundary conditions, we find unacceptably long
autocorrelations for the charge in large subvolumes independent of the boundary
conditions. The long autocorrelations combined with the limited length of our
runs unfortunately prevents us from making a direct numerical comparison of
autocorrelation times between open and periodic boundary conditions. We do find
some success in increasing the rate of topological tunneling by introducing a
dislocation-enhancing determinant, although so far the autocorrelations are
still longer than we would like.


\begin{thebibliography}{99}

\bibitem{Alpha} 
  S.~Schaefer {\it et al.}  [ALPHA Collaboration],
  ``Critical slowing down and error analysis in lattice QCD simulations,''
  Nucl.\ Phys.\ B {\bf 845}, 93 (2011)
  [arXiv:1009.5228 [hep-lat]].

\bibitem{SLOBC} 
  M.~L\"{u}scher and S.~Schaefer,
  ``Lattice QCD without topology barriers,''
  JHEP {\bf 1107}, 036 (2011)
  [arXiv:1105.4749 [hep-lat]].

\bibitem{IwasakiSpacing} 
  A.~Ali Khan {\it et al.}  [CP-PACS Collaboration],
  ``Kaon B parameter from quenched domain wall QCD,''
  Phys.\ Rev.\ D {\bf 64}, 114506 (2001)
  [hep-lat/0105020].

\bibitem{WilsonSpacing} 
  S.~Necco and R.~Sommer,
  ``The $N_f = 0$ heavy quark potential from short to intermediate distances,''
  Nucl.\ Phys.\ B {\bf 622}, 328 (2002)
  [hep-lat/0108008].

\bibitem{5Li} 
  P.~de Forcrand, M.~Garcia Perez and I.-O.~Stamatescu,
  ``Topology of the SU(2) vacuum: A lattice study using improved cooling,''
  Nucl.\ Phys.\ B {\bf 499}, 409 (1997)
  [hep-lat/9701012].

\bibitem{APEradius} 
  C.~W.~Bernard and T.~A.~DeGrand,
  ``Perturbation theory for fat link fermion actions,''
  Nucl.\ Phys.\ Proc.\ Suppl.\  {\bf 83}, 845 (2000)
  [hep-lat/9909083].

\bibitem{OverlapIndex} 
  R.~Narayanan and H.~Neuberger,
  ``Chiral determinant as an overlap of two vacua,''
  Nucl.\ Phys.\ B {\bf 412}, 574 (1994)
  [hep-lat/9307006].

\bibitem{DSDR} 
  R.~Arthur {\it et al.}  [RBC and UKQCD Collaborations],
  ``Domain wall QCD with near-physical pions,''
  Phys.\ Rev.\ D {\bf 87}, 094514 (2013)
  [arXiv:1208.4412 [hep-lat]].

\bibitem{WilsonFlow}
  M.~L\"{u}scher,
  ``Properties and uses of the Wilson flow in lattice QCD,''
  JHEP {\bf 1008}, 071 (2010)
  [arXiv:1006.4518 [hep-lat]].

%\bibitem{MadrasSokal} 
  %N.~Madras and A.~D.~Sokal,
  %``The pivot algorithm: a highly efficient Monte Carlo method for the self-avoiding walk,''
  %J.\ Statist.\ Phys.\  {\bf 50}, 109 (1988).


\end{thebibliography}
\end{document}